\documentclass[aps,prc,preprint,superscriptaddress,showpacs]{revtex4}
\usepackage{amsmath}
\usepackage{amssymb}
\usepackage{bm}
\usepackage{epsfig, graphicx, euscript}
\newcommand{\be}{\begin{equation}}
\newcommand{\ee}{\end{equation}}
\newcommand{\bea}{\begin{eqnarray}}
\newcommand{\eea}{\end{eqnarray}}

\newcommand{\bfr}{\mbox{\boldmath $r$}}
\newcommand{\mbss}[1]{_{\mbox{\scriptsize #1}}}
\newcommand{\mbts}[1]{_{\mbox{\tiny #1}}}
\newcommand{\mbsu}[1]{\mbox{\scriptsize #1}}

\newcommand{\ds}{\displaystyle}

\newcommand{\txts}{\textstyle}

\newcommand{\vk}{\varkappa}

\newcommand{\mbold}[1]{\mbox{\boldmath $#1$}}
\newcommand{\bnabla}{\mbold{\nabla}}
\begin{document}
\title{Description of the Giant Monopole Resonance in the Even-$A$
$^{112-124}$Sn Isotopes within the Microscopic Model Including
Quasiparticle-Phonon Coupling}

\author{A. Avdeenkov}
\affiliation{Institut f\"ur Kernphysik, Forschungszentrum
J\"ulich, 52425 J\"ulich, Germany}
\affiliation{Skobeltsyn Institute of Nuclear Physics,
Moscow State University, 119991 Moscow, Russia}
\author{F. Gr\"ummer}
\affiliation{Institut f\"ur Kernphysik, Forschungszentrum
J\"ulich, 52425 J\"ulich, Germany}
\author{S. Kamerdzhiev}
\affiliation{Institut f\"ur Kernphysik, Forschungszentrum
J\"ulich, 52425 J\"ulich, Germany}
\affiliation{Institute of Physics and Power Engineering,
249033 Obninsk, Russia}
\author{S. Krewald}
\affiliation{Institut f\"ur Kernphysik, Forschungszentrum
J\"ulich, 52425 J\"ulich, Germany}
\author{E. Litvinova}
\affiliation{Gesellschaft f\"{u}r Schwerionenforschung mbH,
Planckstra{\ss }e 1, 64291 Darmstadt, Germany}
\affiliation{Institute of Physics and Power Engineering,
249033 Obninsk, Russia}
\author{N. Lyutorovich}
\affiliation{Institut f\"ur Kernphysik, Forschungszentrum
J\"ulich, 52425 J\"ulich, Germany}
\affiliation{Nuclear Physics Department,
V. A. Fock Institute of Physics, St. Petersburg State University,
198504 St. Petersburg, Russia}
\author{J. Speth}
\affiliation{Institut f\"ur Kernphysik, Forschungszentrum
J\"ulich, 52425 J\"ulich, Germany}
\author{V. Tselyaev}
\affiliation{Institut f\"ur Kernphysik, Forschungszentrum
J\"ulich, 52425 J\"ulich, Germany}
\affiliation{Nuclear Physics Department,
V. A. Fock Institute of Physics, St. Petersburg State University,
198504 St. Petersburg, Russia}
\date{\today}
\begin{abstract}
We have calculated the strength distributions of the giant monopole resonance
in the even-$A$ tin isotopes ($A = 112-124$) which were recently measured
in inelastic $\alpha$-scattering.
The calculations were performed within two microscopic models:
the quasiparticle random phase approximation (QRPA) and
the quasiparticle time blocking approximation which is an extension
of the QRPA including quasiparticle-phonon coupling.
We used a self-consistent calculational scheme based on the HF+BCS
approximation. The single-particle continuum was exactly included
on the RPA level.
The self-consistent mean field and the effective interaction were
derived from the Skyrme energy functional.
In the calculations, two Skyrme force parametrizations were used.
The T5 parametrization with comparatively low value of the
incompressibility of infinite nuclear matter ($K_{\infty}=202$ MeV)
gives theoretical results in good agreement with the experimental data
including the resonance widths.
\end{abstract}
\maketitle

\section{Introduction}

The investigation of the isoscalar giant monopole resonance (ISGMR),
the so-called breathing mode,
is one of the fundamental problems of nuclear physics.
The energy of the ISGMR enables one to determine parameters
characterizing the incompressibility of infinite nuclear matter (INM),
in particular, the value of $K_{\infty}$.
These collective resonances can be studied
experimentally in inelastic $\alpha$-scattering at small angles
(see, e.g., Ref.~\cite{YCL99} and references therein).
Theoretical investigations of these states are based mainly on
the self-consistent microscopic approaches (see, e.g., Refs.
\cite{BGG76,BLM79,B80,TKBM81,BBDG95,FPT97,Giai01,PVCD,STV04,CGMBB,SYZGZ})
including, first of all, scaling and constrained Hartree-Fock (HF) models
and the random phase approximation (RPA)
and on the Landau-Migdal approach that starts with a phenomenological
single-particle basis and with the independently parametrized
particle-hole zero-range interaction
(see, e.g., Refs. \cite{RS74,WM77,KST04} and references therein).
It is important to note that the incompressibility $K_{\infty}$
can not be measured directly but it can be deduced theoretically
by comparing the experimental energies of the ISGMR with
the corresponding calculated values.
The most widely used approach is based on the self-consistent
HF or RPA calculations of the mean energies of the ISGMR
using effective Skyrme or Gogny forces.
Because $K_{\infty}$ can be calculated from the known parameters
of the given force, its value is estimated as the one corresponding
to the force that gives the best description of the experimental
data.
The non-relativistic estimates obtained
in such a way lead to the value $K_{\infty} = 210 \pm 30$ MeV
(see, e.g., Refs. \cite{BGG76,B80,TKBM81,BBDG95,FPT97,Giai01,PVCD,STV04}),
though the recent results favor the upper limit of this estimate
(see \cite{CGMBB,SYZGZ}).
In the Landau-Migdal approach one obtains $K_{\infty}$ from
the scalar-isoscalar Landau-Migdal parameter $f_0$. Here $K_{\infty}$
was always of the order of 240 MeV \cite{RS74}.

Note that within the relativistic mean-field (RMF) theory the INM
incompressibility is usually restricted to the interval
$K_{\infty} = 260 \pm 10$ MeV (see, e.g., Ref.~\cite{VNR03})
that is considerably higher than the non-relativistic limits.
However, recently it was obtained in Ref.~\cite{NVLR08}
that a zero-range (point-coupling) representation of the effective
nuclear interactions in the RMF framework leads to the reduction
of $K_{\infty}$ up to the value of 230 MeV.

In the present paper we investigate theoretically the new
experimental data \cite{GMRexp} on the strength distributions
of the ISGMR in the even-$A$ tin isotopes ($A = 112-124$).
This is the main goal of our work.
The calculations are performed within the framework of the recently
developed microscopic model that takes into account the effects of
the quasiparticle-phonon coupling (QPC) in addition to the usual
correlations included in the conventional RPA.

The paper is organized as follows.
In Sec.~\ref{smod} the model is described. The particular attention
is paid to the dynamical pairing effects which are important to solve
the problem of the $0^+$ spurious state in the ISGMR calculations
in open-shell nuclei.
In Sec.~\ref{gmrcalc} we describe the details of our calculational scheme
and present the results and their discussion.
Conclusions are drawn in the last section.
Appendix~\ref{append1} contains auxiliary formulas.

\section{The model \label{smod}}

\subsection{General scheme}

Two microscopic models were used in our calculations. The first one is
the well-known quasiparticle RPA (QRPA).
The basic ingredients of this approximation are the nuclear mean field
(including the pairing field operator) and the residual
particle-hole (ph) interaction.
In the self-consistent QRPA these ingredients are related to each other
by the consistency condition.
The nuclei excitations are treated as superpositions of
the two-quasiparticle (2q) configurations. This model is applicable to a
wide range of nuclei including open-shell ones as the pairing correlations
of nucleons are taken into account.
The QRPA reproduces well the centroid energies and total strengths
of giant multipole resonances but not their widths. In order
to reproduce the total widths of the resonances it is necessary to enlarge
the configuration space by adding 4q configurations,
i.e. to extend the (Q)RPA.
The most successful approaches in this direction are the models which
take into account the QPC in addition to the correlations included in
the (Q)RPA (see Refs. \cite{S92,BB81,KST04} and references therein).

In the present investigation the QPC contributions
are included within the framework of the recently developed
quasiparticle time blocking approximation (QTBA)
which is an extension of the QRPA in this sense.
On the other hand, since in the QTBA the pairing correlations are also
included, this model is a generalization of the method
of chronological decoupling of diagrams \cite{T89} which is a base
of the Extended Theory of Finite Fermi Systems \cite{KST04}.
Details of the QTBA model are described in Refs. \cite{QTBA1,QTBA2}.
The basic equation of our approach (both in the QRPA and in the QTBA)
is the equation for the effective response function
$R^{\mbsu{eff}}(\omega)$. In the shorthand notations it reads
(we will follow notations of Ref.~\cite{QTBA2})
\be
R^{\mbsu{eff}}(\omega) = A(\omega) - A(\omega)\,{\cal F}\,
R^{\mbsu{eff}}(\omega)
\label{bse1}
\ee
where $A(\omega)$ is a correlated propagator
and $\cal F$ is an amplitude of the effective residual interaction.
In the case of the QRPA, $A(\omega)$ reduces to the uncorrelated
2q propagator $\tilde{A}(\omega)$.
In the general case including pairing correlations,
the amplitude $\cal F$ can be represented as a sum of two terms
\be
{\cal F} = {\cal F}^{(\mbsu{ph})} + {\cal F}^{(\mbsu{pp})}
\label{fsum}
\ee
where the amplitude ${\cal F}^{(\mbsu{ph})}$ represents interaction
in the ph channel and ${\cal F}^{(\mbsu{pp})}$
includes contributions of the interaction
both in the particle-particle (pp) and in the hole-hole (hh) channels
(in the following for brevity we will use the unified term
pp channel implying also the hh-channel contributions).

Let us emphasize that the general formulas of the QTBA derived
in Ref.~\cite{QTBA1} are valid both in the self-consistent and
in the non-self-consistent approaches. In the present paper,
we use self-consistent calculational scheme based on the
HF and Bardeen-Cooper-Schrieffer (BCS) approximations
(in what follows, we will refer to this scheme as the HF+BCS approximation).
The self-consistent mean field and the effective residual interaction were
derived from the Skyrme energy functional
by means of the known variational equations.
In the calculations, the T5 and T6 Skyrme forces
(see Ref.~\cite{TBFP}) were used.

An important property of these parametrizations is that they produce
the nucleon effective mass $m^*$ equal to the bare nucleon mass $m$.
This is a consequence of the fact that the T5 and T6 Skyrme-force parameters
are constrained by the relations (see, e.g., \cite{APDT}):
\bea
t_2 = - \frac{1}{3}\,t_1\,(5 + 4 x_1)\,,\qquad
x_2 = - \frac{4 + 5\,x_1}{5 + 4\,x_1}\,.
\label{ctx2}
\eea
In this case
the contribution of the velocity-dependent terms (except for the
spin-orbital ones) into the energy functional and
the mean field reduces to the derivatives of the nucleon
density, i.e. to the simple surface terms. As a consequence,
contribution of these terms into the effective interaction
derived from such an energy functional also has very simple form.
To see this, consider the energy density $\cal H$
of the Skyrme energy functional $\cal E$ defined as
\be
{\cal E} = \int d \mbold{r}\,{\cal H}(\mbold{r})\,.
\label{enfun}
\ee
In the sufficiently general case it is given, e.g., in Ref.~\cite{TBFP}.
However, if the equations (\ref{ctx2}) hold,
the energy density acquires the form
\bea
{\cal H} &=& \frac{\hbar^2}{2m}(\tau_n + \tau_p)
+ {\txts \frac{1}{2}}\,t_0\,\bigl[ (1 + {\txts \frac{1}{2}}\,x_0)\,\rho^2
- (x_0 + {\txts \frac{1}{2}}) (\rho^2_n + \rho^2_p) \bigr]
\nonumber\\
&+& {\txts \frac{1}{16}}\,t_1\,\bigl\{ 3 (1 + {\txts \frac{1}{2}}\,x_1)
(\bnabla \rho)^2
- 3(x_1 + {\txts \frac{1}{2}})
\bigl[(\bnabla \rho_n)^2 + (\bnabla \rho_p)^2 \bigr]
- x_1 \mbold{J}^2 + \mbold{J}^2_n + \mbold{J}^2_p \bigr\}
\nonumber\\
&-& {\txts \frac{1}{16}}\,t_2\,\bigl\{ (1 + {\txts \frac{1}{2}}\,x_2)
(\bnabla \rho)^2
+ (x_2 + {\txts \frac{1}{2}})
\bigl[(\bnabla \rho_n)^2 + (\bnabla \rho_p)^2 \bigr]
+ x_2 \mbold{J}^2 + \mbold{J}^2_n + \mbold{J}^2_p \bigr\}
\nonumber\\
&+& {\txts \frac{1}{12}}\,t_3\,\bigl[ (1 + {\txts \frac{1}{2}}\,x_3)\,\rho^2
- (x_3 + {\txts \frac{1}{2}}) (\rho^2_n + \rho^2_p) \bigr]\rho^{\alpha}
\nonumber\\
&+& {\txts \frac{1}{2}}\,W_0\,\bigl( \mbold{J} \cdot \bnabla \rho
+ \mbold{J}_n \cdot \bnabla \rho_n + \mbold{J}_p \cdot \bnabla \rho_p \bigr)
+ {\cal H}_{\mbss{Coul}} + {\cal H}_{\mbss{pair}}
\label{enden}
\eea
where ${\cal H}_{\mbss{Coul}}$ is the Coulomb energy density including
the exchange part in the Slater approximation, i.e.
\be
{\cal H}_{\mbss{Coul}}(\mbold{r}) =
\frac{e^2}{2} \int d \mbold{r}'\,
\frac{\rho_p(\mbold{r})\,\rho_p(\mbold{r}')}{|\mbold{r} - \mbold{r}'|}
- \frac{3}{4}\left(\frac{3}{\pi}\right)^{1/3} e^2\,
\rho^{4/3}_p(\mbold{r})\,,
\label{coulen}
\ee
and ${\cal H}_{\mbss{pair}}$ is the density of the pairing energy.
In the applications of the models based on the Skyrme energy functionals
it is frequently taken in the simplest form
\be
{\cal H}_{\mbss{pair}} = {\txts \frac{1}{4}}\,V_0\,
\bigl( \vk^*_n \vk^{\vphantom{*}}_n + \vk^*_p \vk^{\vphantom{*}}_p
\bigr)
\label{pairen}
\ee
which was also used in our calculations.
In Eqs. (\ref{enden})--(\ref{pairen}), $\rho = \rho_n + \rho_p$,
$\rho_q$, $\tau_q$, and $\mbold{J}_q$
are the normal densities and $\vk_q$ is the anomalous
local density of the nucleons of the type $q=n,p$ (neutrons or protons).
In particular, $\rho_q$ is the local particle density,
$\tau_q$ is the kinetic-energy density, and $\mbold{J}_q$
is the spin density. They are defined in the usual way (see, e.g.,
Ref.~\cite{KSTV}). In the case of the spherically symmetric nucleus
and within the HF+BCS approximation they have the form
\bea
\rho^{\vphantom{*}}_q (r) &=& \sum_{(1)}
\delta_{q^{\vphantom{A}}_1,\,q}
\frac{2j^{\vphantom{*}}_1+1}{4\pi}\,
v^2_{(1)}\,R^2_{(1)}(r)\,,
\label{defrho}\\
\tau^{\vphantom{*}}_q (r) &=& \sum_{(1)}
\delta_{q^{\vphantom{A}}_1,\,q}
\frac{2j^{\vphantom{*}}_1+1}{4\pi}\,v^2_{(1)}\,
\bigl[(R^{\,\prime}_{(1)}(r))^2 +
\frac{l^{\vphantom{*}}_1(l^{\vphantom{*}}_1+1)}
{r^2}R^2_{(1)}(r)\bigr]\,,
\label{deftau}\\
\mbold{J}^{\vphantom{*}}_q (r) &=& \frac{\mbold{r}}{r^2}
\sum_{(1)} \delta_{q^{\vphantom{A}}_1,\,q}
\frac{2j^{\vphantom{*}}_1+1}{4\pi}\,v^2_{(1)}\,
\bigl[j^{\vphantom{*}}_1(j^{\vphantom{*}}_1+1) -
l^{\vphantom{*}}_1(l^{\vphantom{*}}_1+1) -
{\txts \frac{3}{4}}\bigr]\,R^2_{(1)}(r)\,,
\label{defspd}\\
\vk^{\vphantom{*}}_q (r) &=& \sum_{(1)}
\delta_{q^{\vphantom{A}}_1,\,q}
\frac{2j^{\vphantom{*}}_1+1}{4\pi}\,
u^{\vphantom{*}}_{(1)}\,v^{\vphantom{*}}_{(1)}\,
R^2_{(1)}(r)\,.
\label{defkap}
\eea
Here and in the following we use the notations of Refs.~\cite{QTBA1,QTBA2}
for the single-quasiparticle basis functions in the doubled space
$\tilde{\psi}^{\vphantom{*}}_1$ which are labelled by the composite indices
$1 = \{[1],m^{\vphantom{*}}_1\}$
where
$[1] = \{(1),\eta^{\vphantom{*}}_1\}$,
$(1)=\{q^{\vphantom{*}}_1,n^{\vphantom{*}}_1,l^{\vphantom{*}}_1,
j^{\vphantom{*}}_1\}$, and $\eta^{\vphantom{*}}_1 = \pm 1$
is the sign of the quasiparticle energy
$E^{\vphantom{*}}_1 = \eta^{\vphantom{*}}_1 E^{\vphantom{*}}_{(1)}$.
I.e., the symbol $(1)$ stands for the set of the single-particle
quantum numbers excepting the projection of the total angular momentum
$m^{\vphantom{*}}_1$,
$R^{\vphantom{*}}_{(1)}(r)$ is the radial part of the
single-particle wave function, $v^2_{(1)}$ is the occupation
probability, and $u^{\vphantom{*}}_{(1)}=\sqrt{1-v^2_{(1)}}$.

As can be seen from Eqs. (\ref{enfun}) and (\ref{enden}),
the following equality holds
\be
\frac{\delta {\cal E}}{\delta \tau_q(\bfr)} =
\frac{\hbar^2}{2m} = \mbox{constant}\,.
\label{d1tau}
\ee
In particular, it means that
the equations of motion derived from such an energy functional $\cal E$
contain the nucleon effective mass $m^*_q(r)=m$.
The spin-scalar part of the effective interaction in the
ph channel corresponding to $\cal E$
is determined by the relation
\be
{\cal F}^{(\mbsu{ph})}_{0,\,qq'}(\bfr,\bfr') = \frac{\delta^2 {\cal E}}
{\delta \rho_q(\bfr)\,\delta \rho_{q'}(\bfr')}\,.
\label{defint}
\ee
This ansatz completely includes velocity-dependent contributions
because of Eq.~(\ref{d1tau}). In the explicit form we have
\bea
{\cal F}^{(\mbsu{ph})}_{0,\,nn}(\bfr,\bfr') &=&
\biggl( {\txts \frac{1}{2}}\,t_0\,(1-x_0) +
{\txts \frac{1}{12}}\,t_3\,\biggl\{ (1 + {\txts \frac{1}{2}}\,x_3)\,
(1 + \alpha)\,(2 + \alpha)\,\rho^{\alpha}
\nonumber\\
&-& (x_3 + {\txts \frac{1}{2}})\,\bigl[ 2\rho^{\alpha}
+4\alpha \rho^{\vphantom{2}}_n \rho^{\alpha -1}
+\alpha (\alpha -1) (\rho^2_n + \rho^2_p)\rho^{\alpha -2} \bigr] \biggr\}
\biggr)\,\delta (\bfr - \bfr')
\nonumber\\
&+& {\txts \frac{3}{16}}\,\bigl[ t_2\,(1+x_2) - t_1\,(1-x_1) \bigr]
\Delta\,\delta (\bfr - \bfr')\,,
\label{fphnn}
\eea
\bea
{\cal F}^{(\mbsu{ph})}_{0,\,np}(\bfr,\bfr') &=&
\biggl( t_0\,(1+{\txts \frac{1}{2}}\,x_0) +
{\txts \frac{1}{12}}\,t_3\,\biggl\{ (1 + {\txts \frac{1}{2}}\,x_3)\,
(1 + \alpha)\,(2 + \alpha)\,\rho^{\alpha}
\nonumber\\
&-& (x_3 + {\txts \frac{1}{2}})\,\alpha \,\bigl[ (\alpha +1)\rho^{\alpha}
-2(\alpha -1) \rho^{\vphantom{2}}_n\,\rho^{\vphantom{2}}_p\,\rho^{\alpha -2}
\bigr] \biggr\} \biggr)\,\delta (\bfr - \bfr')
\nonumber\\
&+& {\txts \frac{1}{8}}\,\bigl[ t_2\,(1+{\txts \frac{1}{2}}x_2)
- 3t_1\,(1+{\txts \frac{1}{2}}x_1) \bigr]
\Delta\,\delta (\bfr - \bfr')\,.
\label{fphnp}
\eea
The formulas for the components ${\cal F}^{(\mbsu{ph})}_{0,\,pp}$ and
${\cal F}^{(\mbsu{ph})}_{0,\,pn}$ are obtained from Eqs. (\ref{fphnn})
and (\ref{fphnp}) by replacing indices $n$ by $p$ and $p$ by $n$
and by adding the Coulomb interaction to ${\cal F}^{(\mbsu{ph})}_{0,\,pp}$.

Let us note that in addition to the simplicity of the formulas for
the effective residual interaction there exist the physical reasons
to use the Skyrme forces with $m^*=m$. It is known that for heavy
and medium mass nuclei the single-particle spectra obtained in the
HF calculations with such Skyrme forces better reproduce the experimental
energies as compared with the case of the forces with $m^*/m \sim 0.7$.
This results in better description of the excitations of the even-mass
nuclei in the RPA and QRPA. The same is true for the QTBA if
the subtraction procedure
(see Eq.~(\ref{bphi}) below and Refs.~\cite{QTBA1,QTBA2}) is used.

The spin-vector components of the effective interaction are not
determined uniquely from Eq.~(\ref{enden}) which is valid only for
the spin-saturated nuclei. In our calculations these components
are taken in the simple form of the Landau-Migdal zero-range force
with known parameters $C_0$, $g$, and $g'$
(see, e.g., Ref.~\cite{QTBA2}).
However, the spin-vector components of the interaction do not
enter equations for the $0^+$ excitations.

Note that in the fully self-consistent HF-RPA calculations \cite{SSAR}
it was obtained that the spin-orbital and the Coulomb components
of the effective residual interaction slightly shift the mean energy
of the ISGMR in the opposite directions. This effect is also
confirmed by our estimates of the mean energies of the ISGMR
based on the constrained HF method described in Ref.~\cite{BLM79}
(see also \cite{STV04} for details of our calculational scheme).
Namely, in the test calculations for $^{120}$Sn we have obtained
the following results using T5 Skyrme force and neglecting pairing
correlations. The constrained HF method (fully including
the spin-orbital and the Coulomb contributions) yields for the mean
energy $\sqrt{m_1/m_{-1}}$ the value 15.1 MeV where $m_1$ and $m_{-1}$
are the energy-weighted moments defined by Eq.~(\ref{defmk}) below
for the infinite energy interval ($E_1=0$, $E_2=\infty$).
For the same interval we obtain in the RPA the value 15.4 MeV if we
neglect only the spin-orbital contributions in the residual interaction.
If we neglect both the spin-orbital and the Coulomb contributions
in the RPA interaction we obtain the value 15.0 MeV
that differs from the constrained HF result only by 0.1 MeV.

In our QRPA and QTBA calculations, the spin-orbital components of the
effective interaction (but not of the mean field) are neglected.
For the reasons described above we also neglect the Coulomb contribution
into ${\cal F}^{(\mbsu{ph})}_{0,\,pp}$.
The effective interaction in the pp channel
and the gap equation within the HF+BCS approximation are determined
by the formulas of Appendix A of Ref.~\cite{QTBA2} with
${\cal F}^{\xi}(r)=\frac{1}{2}V_0$
(see also Appendix~\ref{append1} of the present paper).

\subsection{Dynamical pairing effects in QRPA and QTBA}

One of the important questions arising in the QRPA and QTBA calculations
is the question of completeness of the configuration space.
The size of the basis in this space has an impact practically on all
the calculated quantities. In particular, configurations with a particle
in the continuum are responsible for the formation of the escape widths
of the resonances.
The well-known method to include these configurations on the RPA level
is the use of the coordinate representation within the Green function
formalism (see Ref.~\cite{SB75}). We use this method in our approach
as described in Ref.~\cite{QTBA2}.
However, incorporation of the pp-channel contributions in
the coordinate representation leads to considerable numerical
difficulties. At the same time, the pp-channel contributions
(so-called dynamical pairing effects) are very important in the
calculations of $0^+$ excitations in the open-shell nuclei,
first of all because of the problem of the $0^+$ spurious state.
For this reason we have developed a combined method
which is a modification of the so-called $(r,\lambda)$ representation
proposed in Ref.~\cite{PS88} for the QRPA problem.
Within this method only the ph channel is treated in the coordinate space,
while the dynamical pairing effects
are included in the discrete basis representation.

Considering the general case of the QTBA, note that
taking into account decomposition (\ref{fsum}) one can rewrite
Eq.~(\ref{bse1}) in the form
\be
R^{\mbsu{eff}}(\omega) = A^{(\mbsu{res+pp})}(\omega)
- A^{(\mbsu{res+pp})}(\omega)\,{\cal F}^{(\mbsu{ph})}\,
R^{\mbsu{eff}}(\omega)
\label{bse2}
\ee
where propagator $A^{(\mbsu{res+pp})}(\omega)$ is a solution
of the equation
\be
A^{(\mbsu{res+pp})}(\omega) = A(\omega)
- A(\omega)\,{\cal F}^{(\mbsu{pp})}\,
A^{(\mbsu{res+pp})}(\omega)\,.
\label{app1}
\ee
In the present work we use the version of the QTBA in which
the ground state correlations caused by the QPC are neglected.
In this case the correlated propagator $A(\omega)$ is defined
by the equation
\be
A(\omega) = \tilde{A}(\omega)
- \tilde{A}(\omega)\,\bar{\Phi}(\omega)\,A(\omega)
\label{cprp}
\ee
where $\tilde{A}(\omega)$ is the uncorrelated QRPA propagator,
\be
\bar{\Phi}(\omega) =
\Phi^{(\mbsu{res})}(\omega) - \Phi^{(\mbsu{res})}(0)\,,
\label{bphi}
\ee
and $\Phi^{(\mbsu{res})}(\omega)$ is a resonant part of the
interaction amplitude responsible for the QPC in our model
(see Refs.~\cite{QTBA2,QTBA1} for details).
Combination of Eqs. (\ref{app1}) and (\ref{cprp}) leads to the
new equation for $A^{(\mbsu{res+pp})}(\omega)$:
\be
A^{(\mbsu{res+pp})}(\omega) = \tilde{A}(\omega) - \tilde{A}(\omega)
\bigl[\bar{\Phi}(\omega) + {\cal F}^{(\mbsu{pp})}\bigr]
A^{(\mbsu{res+pp})}(\omega)\,.
\label{app2}
\ee

As a result we obtain that the pp-channel contributions can be included
by modification of the equation for the correlated propagator, i.e.
by replacing Eq.~(\ref{cprp}) by Eq.~(\ref{app2}). The modification
is reduced to the additional term ${\cal F}^{(\mbsu{pp})}$ added to
the amplitude $\bar{\Phi}(\omega)$.
The respective equations in terms of the reduced matrix elements
are drawn in Appendix~\ref{append1}.
It is worth noting that the QPC in the QTBA
is included both in the ph channel and in the pp channel because
there is no difference between these channels in the representation
of the single-quasiparticle basis functions in the doubled space
($\tilde{\psi}_1$, see Ref.~\cite{QTBA2})
which is used in Eqs. (\ref{app1}), (\ref{cprp}), and
(\ref{app2}). This is true both for the system of equations
(\ref{bse1}), (\ref{cprp}) and for the system (\ref{bse2}),
(\ref{app2}).

Note, however, that in practice Eq.~(\ref{bse2}) for
$R^{\mbsu{eff}}(\omega)$ is solved in the coordinate representation
(to take into account the single-particle continuum),
while Eq.~(\ref{app2})
is solved in the restricted discrete basis representation.
This fact greatly simplifies the problem as compared with the initial
Eq.~(\ref{bse1}) in which both the ph-channel contribution and
the pp-channel one are included in the coordinate representation.
On the other hand, the use of the restricted discrete basis
representation for the pp channel is fully consistent with
BCS approximation in which the gap equation is solved in the same
restricted basis.

The general scheme described above ensures that the energy of the
$0^+$ spurious state (so-called ghost state) is equal to zero
both in the QRPA and in the QTBA.
However, there still remains the following problem:
in the QTBA the ghost state can be fragmented due to its coupling
to the 2q$\otimes$phonon configurations,
despite the energy of the dominant ghost state is equal to zero.
It can lead to the spurious states at low energies distorting respective
strength functions. In particular, these fragmented spurious states
will produce non-zero response to the particle-number operator
which has to be exactly equal to zero in a correct theory
(as, for instance, in the QRPA including pp channel that was proved
by Migdal, see \cite{M67}).
In the present calculations this problem is solved with the help
of special projection technique which will be described in
a forthcoming publication.

\section{Calculations of the Giant Monopole Resonance
in the tin isotopes \label{gmrcalc}}

\subsection{Numerical details}

The method described above has been applied to calculate
the strength distributions of the isoscalar giant monopole resonance
in the even-$A$ tin isotopes ($A = 112-124$) which were recently measured
experimentally with inelastic scattering of $\alpha$ particles
(see Ref.~\cite{GMRexp}).
The ground state properties of these nuclei were calculated within
HF+BCS approximation using T5 and T6 Skyrme forces with
the parameters drawn in Ref.~\cite{TBFP} including
the pairing-force strength $V_0 = -210$ MeV$\cdot$fm$^3$
in Eq.~(\ref{pairen}).
For all tin isotopes under consideration,
the pairing window for the neutrons contains 22 states including
all the discrete states and one or two quasidiscrete states.
The criterion to select quasidiscrete states is described
in Ref.~\cite{QTBA2}.

To calculate the strength function of the ISGMR, the equation (\ref{bse1})
for the effective response function $R^{\mbsu{eff}}(\omega)$
was solved. The strength function $S(E)$ is determined by
$R^{\mbsu{eff}}(\omega)$ via the formula
\be
S(E) = \frac{1}{2\pi}\,\mbox{Im}\sum_{1234}\,
(eV^{\,0})^{\ds *}_{21}
\,R^{\,\mbsu{eff}}_{12,34} (E + i \Delta)\,
(eV^{\,0})^{\vphantom{\ds *}}_{43}
\label{defsf}
\ee
where $E$ is an excitation energy,
$\Delta$ is a smearing parameter, $V^0$ is an external field,
and $e$ is an effective charge operator.
In the case of the isoscalar $0^+$ excitations the one-body operator
$eV^0$ is proportional to the identity matrices
both in the spin and in the isospin indices.
Its radial dependence is taken in our calculations in the form
$eV^0=r^2$. The smearing parameter was taken to be equal to 500~keV
that approximately corresponds to the experimental resolution
for the data presented in Ref.~\cite{GMRexp}.

In the calculation of the QTBA correlated propagator $A(\omega)$
entering Eq.~(\ref{bse1}),
the valence zone for the neutrons coincides with the pairing window.
The valence zone for the protons contains 20 states including
all the discrete states and several quasidiscrete states
as described in Ref.~\cite{QTBA2}.
Let us emphasize that the restricted valence zone is used only
in the calculation of the discrete part of the propagator $A(\omega)$
including QPC effects and in the calculation of the phonons (see below).
In the ISGMR calculations, the configurations with the particle in the
continuum are included completely in the RPA-like part of $A(\omega)$
(see Ref.~\cite{QTBA2} for details).

The set of phonons in the QTBA calculations included collective modes
with values of the spin $L$ in the interval $2\leqslant L \leqslant 9$
and with natural parity $\pi = (-1)^L$.
The phonon characteristics were calculated within the QRPA
using configuration space restricted by the valence zone described above.
The maximal energy of the
phonon was adopted to be equal to the value 10 MeV which is approximately
equal to the nucleon separation energy for the given tin isotopes.
The second criterion to include the phonon into the phonon space was
its reduced transition probability $B(EL)$ which should be more than
10\% of the maximal $B(EL)$ for the given spin.
According to these criterions, the total number of phonons included
in the QTBA calculations is equal to
21 for $^{112}$Sn,
19 for $^{114}$Sn,
23 for $^{116}$Sn,
26 for $^{118}$Sn,
29 for $^{120}$Sn,
27 for $^{122}$Sn, and
31 for $^{124}$Sn.

To describe correctly effects of a fragmentation of the resonances
in the QTBA arising due to the QPC it is very important to use
the phonon space with the phonon characteristics close to the
experimental ones.
However, both the T5 and the T6 Skyrme forces do not provide
satisfactory description of the experimental energies and transition
probabilities within the self-consistent QRPA scheme
presented in Section~\ref{smod}. For this reason, in the calculation
of the phonons (and only in this calculation)
we have used the QRPA scheme which is self-consistent
only on the mean-field level.
More specifically, the mean field was calculated within HF+BCS
approximation based on the T5 Skyrme force,
while the effective residual interaction was taken in the form of
the Landau-Migdal zero-range force with the standard set of the
parameters (see, e.g., Ref.~\cite{KLLT}), except for the parameter
$f_{\mbss{ex}}$.
This parameter was adjusted for the each nucleus to reproduce the
experimental energies of the $2^{+}_1$ and $3^{-}_1$ levels.
As a result, the parameter $f_{\mbss{ex}}$ takes the values
in the interval $-1.54 \pm 0.11$ for the phonons
with the positive parity and the values
in the interval $-1.83 \pm 0.06$ for the phonons
with the negative parity.

\subsection{Results and discussion}

\begin{figure}
\includegraphics*[scale=0.81,angle=0]{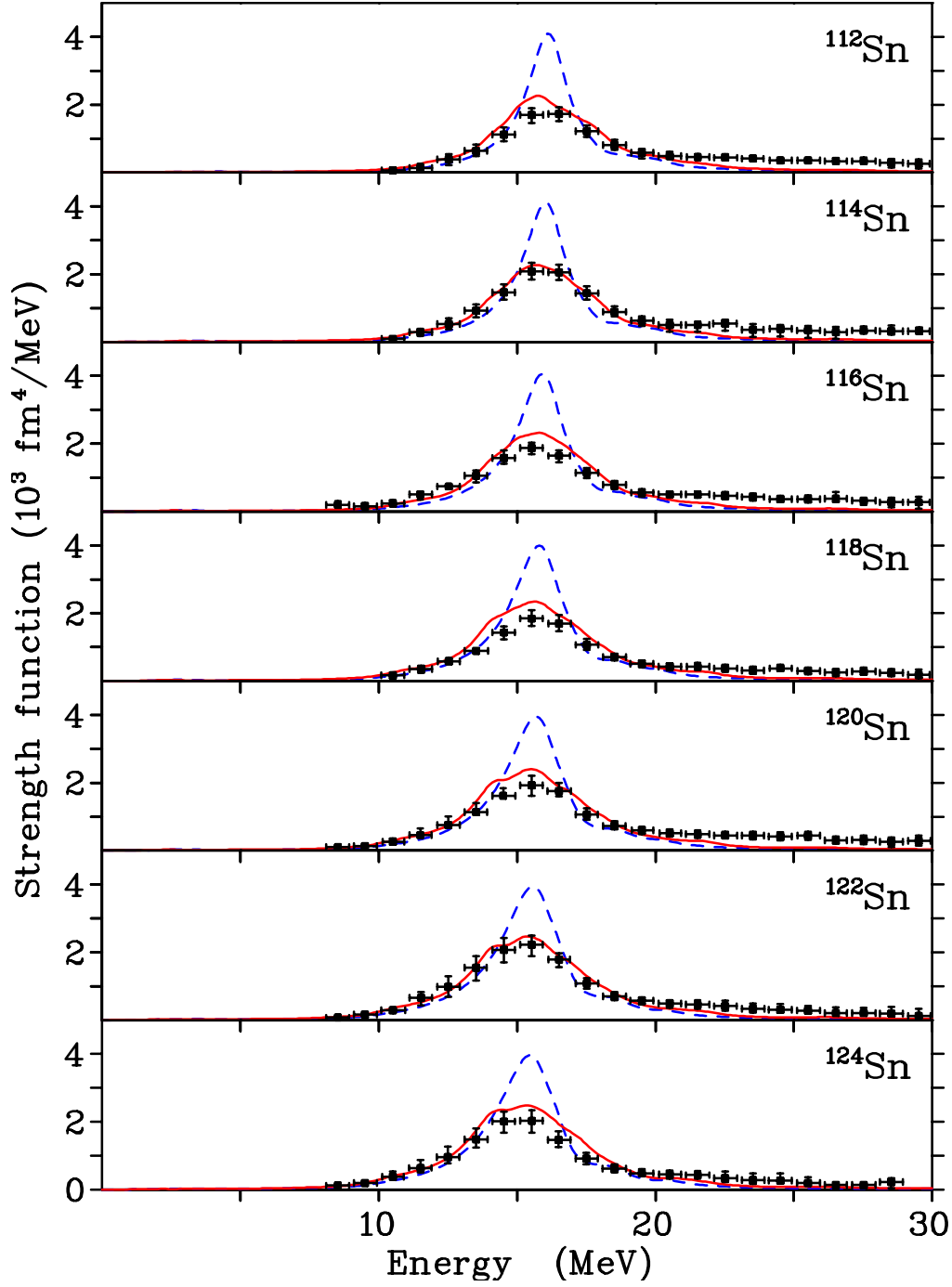}
\caption{\label{sn7is}
Isoscalar giant monopole resonance in the even-$A$ $^{112-124}$Sn isotopes
calculated within QRPA (dashed line) and QTBA (solid line).
The results are obtained
within self-consistent HF+BCS approach based on the T5 Skyrme force.
The smearing parameter $\Delta$ is equal to 500 keV.
Experimental data (solid squares) are taken from Ref.~\cite{GMRexp}.}
\end{figure}

The results for the ISGMR strength distributions in the even-$A$
$^{112-124}$Sn isotopes are presented in Fig.~\ref{sn7is}
and in Table~\ref{tab1}.
The energy-weighted moments $m_k$ were determined as
\be
m_k = \int^{E_2}_{E_1} E^k S(E)\,d E\,.
\label{defmk}
\ee

\noindent
The energy interval limited by $E_1=10.5$ MeV and $E_2=20.5$ MeV
was taken the same as in Ref.~\cite{GMRexp}.
As can be seen from the Table~\ref{tab1},
the agreement of the theoretical results with the experimental
mean energies is good both in the QRPA and in the QTBA.
The fact that the mean energies obtained in the QRPA and in the QTBA
are very close to each other is explained by the subtraction procedure
used in our calculations (see Eq.~(\ref{bphi}) and Ref.~\cite{QTBA2}
for the discussion).
The main reason of the agreement with the experiment is that
in this calculation we used the self-consistent scheme
based on the T5 Skyrme-force
parametrization with comparatively low value of the
incompressibility of INM ($K_{\infty}=202$ MeV).
The other parametrizations with
$K_{\infty}$ around 240 MeV give too large mean energies of the ISGMR
in the considered tin isotopes as compared with the experiment.

\begin{table}
\caption{\label{tab1}
Mean energies and widths for the ISGMR
strength distributions in the even-$A$ $^{112-124}$Sn isotopes calculated
for 10.5--20.5 MeV energy interval. Theoretical results are obtained
within self-consistent HF+BCS approach based on the T5 Skyrme force
($K_{\infty}=202$ MeV).
Experimental values are taken from Ref.~\cite{GMRexp}.
The RPA results for $^{100,132}$Sn and $^{208}$Pb are drawn for comparison.
The values for $^{208}$Pb are calculated for 5--25 MeV energy interval.}
\begin{ruledtabular}
\begin{tabular}{clcccc}
&& $\sqrt{m_1/m_{-1}}$, MeV & $m_1/m_0$, MeV
 & $\sqrt{m_3/m_1}$, MeV & $\Gamma$, MeV \\
\hline
$^{100}$Sn & RPA        & 16.4 & 16.5 & 16.7 & 2.2 \\
\hline
           & QRPA       & 16.0 & 16.1 & 16.3 & 1.9 \\
$^{112}$Sn & QTBA       & 15.9 & 16.0 & 16.4 & 3.9 \\
           & Experiment & 16.1 $\pm$ 0.1 & 16.2 $\pm$ 0.1 & 16.7 $\pm$ 0.2
           & 4.0 $\pm$ 0.4 \\
\hline
           & QRPA       & 15.9 & 15.9 & 16.2 & 2.0 \\
$^{114}$Sn & QTBA       & 15.7 & 15.9 & 16.3 & 3.9 \\
           & Experiment & 15.9 $\pm$ 0.1 & 16.1 $\pm$ 0.1 & 16.5 $\pm$ 0.2
           & 4.1 $\pm$ 0.4 \\
\hline
           & QRPA       & 15.7 & 15.8 & 16.1 & 2.1 \\
$^{116}$Sn & QTBA       & 15.6 & 15.7 & 16.1 & 4.0 \\
           & Experiment & 15.7 $\pm$ 0.1 & 15.8 $\pm$ 0.1 & 16.3 $\pm$ 0.2
           & 4.1 $\pm$ 0.3 \\
\hline
           & QRPA       & 15.6 & 15.7 & 16.0 & 2.3 \\
$^{118}$Sn & QTBA       & 15.5 & 15.6 & 16.0 & 4.2 \\
           & Experiment & 15.6 $\pm$ 0.1 & 15.8 $\pm$ 0.1 & 16.3 $\pm$ 0.1
           & 4.3 $\pm$ 0.4 \\
\hline
           & QRPA       & 15.4 & 15.5 & 15.8 & 2.4 \\
$^{120}$Sn & QTBA       & 15.3 & 15.5 & 15.9 & 4.3 \\
           & Experiment & 15.5 $\pm$ 0.1 & 15.7 $\pm$ 0.1 & 16.2 $\pm$ 0.2
           & 4.9 $\pm$ 0.5 \\
\hline
           & QRPA       & 15.3 & 15.4 & 15.7 & 2.5 \\
$^{122}$Sn & QTBA       & 15.2 & 15.4 & 15.8 & 4.4 \\
           & Experiment & 15.2 $\pm$ 0.1 & 15.4 $\pm$ 0.1 & 15.9 $\pm$ 0.2
           & 4.4 $\pm$ 0.4 \\
\hline
           & QRPA       & 15.2 & 15.3 & 15.6 & 2.6 \\
$^{124}$Sn & QTBA       & 15.1 & 15.3 & 15.7 & 4.4 \\
           & Experiment & 15.1 $\pm$ 0.1 & 15.3 $\pm$ 0.1 & 15.8 $\pm$ 0.1
           & 4.5 $\pm$ 0.5 \\
\hline
$^{132}$Sn & RPA        & 14.8 & 14.9 & 15.2 & 2.7 \\
\hline
$^{208}$Pb & RPA        & 12.7 & 12.9 & 13.4 & 1.8 \\
\end{tabular}
\end{ruledtabular}
\end{table}

For comparison, in Table~\ref{tab2} we draw the QRPA results
obtained with the T6 Skyrme force ($K_{\infty}=236$ MeV).
As can be seen, the T6 mean energies $m_1/m_0$ are greater than
the experimental values for the tin isotopes by more than one MeV.
This fact agrees with the results of Ref.~\cite{P07} where
it was obtained that the relativistic RPA calculations
based on the force with $K_{\infty}=230$ MeV consistently overestimate
the centroid energies of the ISGMR in the same tin isotopes.

\begin{table}[!t]
\caption{\label{tab2} Mean energies and widths for the ISGMR
strength distributions in the even-$A$ $^{112-124}$Sn isotopes
calculated within the QRPA for 10.5--20.5 MeV energy interval. The
calculations were performed within self-consistent HF+BCS approach
based on the T6 Skyrme force ($K_{\infty}=236$ MeV).
The RPA results for $^{100,132}$Sn and $^{208}$Pb are also shown.
The values for $^{208}$Pb are calculated for 5--25 MeV energy interval.}
\begin{ruledtabular}
\begin{tabular}{ccccc}
& $\sqrt{m_1/m_{-1}}$, MeV & $m_1/m_0$, MeV
& $\sqrt{m_3/m_1}$, MeV & $\Gamma$, MeV \\
\hline
$^{100}$Sn & 17.4 & 17.5 & 17.8 & 2.4 \\
$^{112}$Sn & 17.1 & 17.2 & 17.4 & 2.2 \\
$^{114}$Sn & 17.0 & 17.1 & 17.3 & 2.3 \\
$^{116}$Sn & 16.8 & 16.9 & 17.2 & 2.4 \\
$^{118}$Sn & 16.7 & 16.8 & 17.1 & 2.5 \\
$^{120}$Sn & 16.6 & 16.7 & 17.0 & 2.7 \\
$^{122}$Sn & 16.5 & 16.6 & 16.9 & 2.8 \\
$^{124}$Sn & 16.3 & 16.5 & 16.8 & 3.0 \\
$^{132}$Sn & 16.0 & 16.1 & 16.5 & 3.2 \\
$^{208}$Pb & 13.9 & 14.1 & 14.7 & 2.0 \\
\end{tabular}
\end{ruledtabular}
\end{table}

Note that the value $K_{\infty}=202$ MeV corresponding to
the T5 Skyrme force lies within the interval
$210 \pm 30$ MeV which was considered for a long time
as the non-relativistic estimate for this quantity.
The recent results \cite{CGMBB,SYZGZ} giving $K_{\infty}$ around
230--240 MeV were obtained on the base of the experimental data
in fact only for the one nucleus $^{208}$Pb which is doubly magic.
However, the question which arises is whether the doubly magic nucleus
is the best candidate to determine the value of the INM incompressibility
in view of the strong shell effects taking place in this case.
On the other hand, by comparing the RPA results for $^{208}$Pb
shown in the Tables \ref{tab1} and \ref{tab2} one can see that
the T6 Skyrme force with $K_{\infty}=236$ MeV
nicely reproduces the experimental data for this
nucleus ($(m_1/m_0)_{\mbss{exp}}$ = 14.2 $\pm$ 0.3 MeV, see \cite{YCL99}),
while the T5 force gives the result which is lesser by 1.3~MeV.
Thus, the question about the precise value of $K_{\infty}$
is not resolved within the framework of our approach.

The theoretical values of the ISGMR widths $\Gamma$ were obtained
from the Lorentzian fit of the calculated functions $S(E)$.
In contrast to the mean energies, the QRPA and the QTBA give substantially
different results for the width. It is well known that the spreading
width $\Gamma^{\downarrow}$ is a considerable part of the total width
of the giant resonance. The QRPA does not produce $\Gamma^{\downarrow}$,
while in the QTBA it is formed by the 2q$\otimes$phonon configurations.
This is the reason why the QRPA strongly underestimates the experimental
values of $\Gamma$, while very good agreement is achieved in the QTBA.

To investigate the nature of dependence of the ISGMR mean energies
on the neutron excess $(N-Z)$ we calculated the unperturbed
$0^+_{\mbss{IS}}$ response
substituting the (Q)RPA uncorrelated propagator $\tilde{A}(\omega)$
in Eq.~(\ref{defsf}) instead of $R^{\,\mbsu{eff}}(\omega)$.
This response corresponds to the independent quasiparticle model (IQM).
The results are presented in Table~\ref{tab3} in comparison with
the (Q)RPA results obtained in the same energy interval 10--30 MeV.
\begin{table}[!t]
\caption{\label{tab3}
Mean energies for the $0^+_{\mbss{IS}}$ strength distributions
in the even-$A$ $^{100,112-124,132}$Sn isotopes calculated for 10--30 MeV
energy interval
within self-consistent HF+BCS approach based on the T5 Skyrme force.
See text for details.}
\begin{ruledtabular}
\begin{tabular}{clccc}
&& $\sqrt{m_1/m_{-1}}$, MeV & $m_1/m_0$, MeV
&  $\sqrt{m_3/m_1}$, MeV  \\
\hline
$^{100}$Sn & IQM  & 19.1 & 19.3 & 19.9 \\
           & RPA  & 16.8 & 17.0 & 17.6 \\
\hline
$^{112}$Sn & IQM  & 18.5 & 18.7 & 19.4 \\
           & QRPA & 16.4 & 16.5 & 17.1 \\
\hline
$^{114}$Sn & IQM  & 18.3 & 18.5 & 19.3 \\
           & QRPA & 16.2 & 16.4 & 17.0 \\
\hline
$^{116}$Sn & IQM  & 18.1 & 18.4 & 19.1 \\
           & QRPA & 16.0 & 16.2 & 16.8 \\
\hline
$^{118}$Sn & IQM  & 17.9 & 18.2 & 19.0 \\
           & QRPA & 15.9 & 16.0 & 16.7 \\
\hline
           & IQM  & 17.8 & 18.1 & 18.9 \\
$^{120}$Sn & QRPA & 15.7 & 15.9 & 16.5 \\
           & RPA  & 15.2 & 15.4 & 16.0 \\
\hline
$^{122}$Sn & IQM  & 17.6 & 17.9 & 18.7 \\
           & QRPA & 15.6 & 15.7 & 16.4 \\
\hline
$^{124}$Sn & IQM  & 17.5 & 17.8 & 18.6 \\
           & QRPA & 15.4 & 15.6 & 16.3 \\
\hline
$^{132}$Sn & IQM  & 17.1 & 17.4 & 18.2 \\
           & RPA  & 15.0 & 15.2 & 15.8 \\
\end{tabular}
\end{ruledtabular}
\end{table}
This interval was chosen to exclude contribution of the low-lying
strength arising in the IQM response.
As can be seen from Table~\ref{tab3}, the $(N-Z)$ dependence
of the (Q)RPA mean energies practically follows the dependence
of the IQM energies. In particular, the difference between the
$m_1/m_0$ values for $^{112}$Sn and $^{124}$Sn in the QRPA
is equal to 0.9~MeV and the same difference is obtained in the
IQM calculation.
Since the poles of the uncorrelated propagator $\tilde{A}(\omega)$
are equal to the sums of the quasiparticle energies
$E_{(1)} + E_{(2)}$ [see Eq.~(\ref{ajpp3})],
this result means that the $(N-Z)$ dependence of the ISGMR mean energies
is mainly determined by the level density of the single-quasiparticle
spectrum.
Including the residual interaction in the (Q)RPA, we obtain
the following redistribution of the isoscalar monopole strength:
the low-lying part of the strength disappears,
while the mean energy of the high-lying states (which form the ISGMR)
is reduced by approximately two MeV.

\begin{figure}[!t]
\includegraphics*[scale=0.65,angle=90]{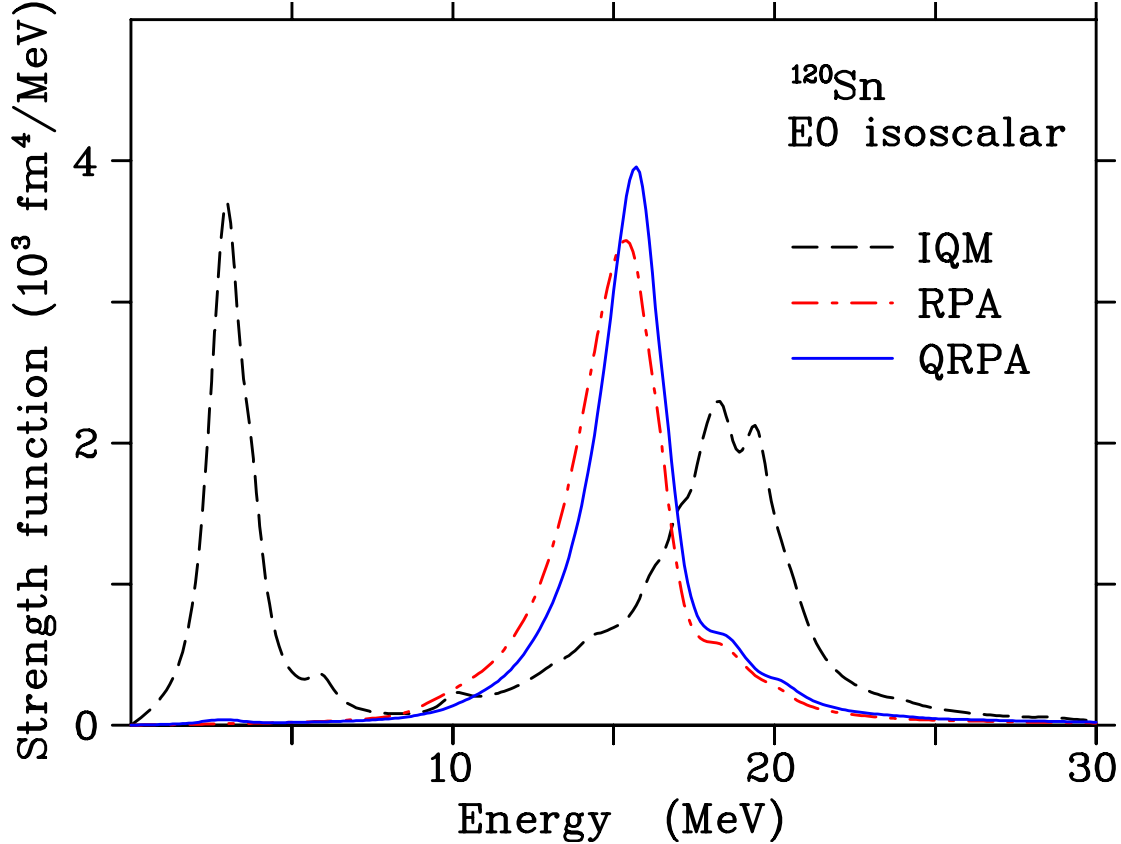}
\caption{\label{sn120iqm}
Isoscalar E0 response in $^{120}$Sn
calculated within the independent quasiparticle model (IQM, dashed line),
RPA (dashed-dotted line), and QRPA (solid line), making use of
the T5 Skyrme force. See text for details.
The smearing parameter $\Delta$ is equal to 500 keV.}
\end{figure}

In Table~\ref{tab3}, we also include the ISGMR mean energies obtained
within the RPA for $^{120}$Sn nucleus. In this calculation, the pairing
correlations are neglected both in the mean field and in the
residual interaction.
Respective strength function is shown in Fig.~\ref{sn120iqm}
in comparison with the IQM and QRPA strength functions.
These results demonstrate that the influence of
the pairing correlations on the ISGMR mean energies is appreciable.
In the QRPA, where the pairing correlations are included,
the mean energies increase by 0.5 MeV as compared with the RPA.
By comparing the results obtained with the T5 and T6 Skyrme forces,
one can see that this increase is substantial for determining
the INM incompressibility.

\section{Conclusions}

In the paper the results of the theoretical analysis of the ISGMR
strength distributions in the even-$A$ $^{112-124}$Sn isotopes are presented.
The calculations were performed within two microscopic models:
the quasiparticle random phase approximation (QRPA) and
the quasiparticle time blocking approximation (QTBA) which is an extension
of the QRPA including quasiparticle-phonon coupling.
We used self-consistent calculational scheme based on the HF+BCS
approximation.
The self-consistent mean field and the effective interaction were
derived from the Skyrme energy functional. In the calculations,
two Skyrme force parametrizations were used. The T5 parametrization
with comparatively low value of the incompressibility
of infinite nuclear matter ($K_{\infty}=202$ MeV)
allowed us to achieve good agreement with the experimental data
for tin isotopes within the QTBA including resonance widths.
However, this parametrization
fails to reproduce the experimental ISGMR energy for the $^{208}$Pb nucleus
which is usually used in the fit of the Skyrme force parameters.
On the other hand, the T6 Skyrme force with $K_{\infty}=236$ MeV
nicely reproduces the ISGMR energy for $^{208}$Pb but overestimates
the energies for $^{112-124}$Sn isotopes by more than one MeV.
On the whole, these results do not allow us
to decrease the ambiguity in the value of
$K_{\infty}$ as compared with the previous known estimates.
Note, however, that the main goal of our work is not to solve
the problem of the nuclear matter incompressibility but to find
under which conditions one can obtain reasonable description
of the experimental data for the considered tin isotopes
within the framework of the self-consistent approach
including correlations beyond the QRPA.

\begin{acknowledgements}
The work was supported by Deutsche Forschungsgemeinschaft
under the grant No. 436 RUS 113/806/0-1
and by the Russian Foundation for Basic Research
under the grant No. 05-02-04005-DFG\_a.
V.~T. thanks the Institut f\"ur Kernphysik at the
Forschungszentrum J\"ulich for hospitality during the
completion of this work.
\end{acknowledgements}

\appendix

\section{Modification of the QTBA equations
including contribution of the particle-particle channel
in terms of the reduced matrix elements \label{append1}}

In the detailed form using the notations of Ref.~\cite{QTBA2}
for the reduced matrix elements, our method
to include pp-channel contribution in the QTBA equations consists
in the following. In Eq.~(33) of Ref.~\cite{QTBA2} only ph channel
is kept, but in Eq.~(42) for
$A^{J\,(\mbsu{ph,ph})\,LS,L'S'}_{[12,34]}(\omega)$ the matrix element
$A^{J}_{[12,34]}(\omega)$ is replaced by
$A^{J\,(\mbsu{res+pp})}_{[12,34]}(\omega)$ where
\be
A^{J\,(\mbsu{res+pp})}_{[12,34]}(\omega) =
\delta^{\vphantom{(+)}}_{\eta_{\mbts{1}},-\eta_{\mbts{2}}}\,
\delta^{\vphantom{(+)}}_{\eta_{\mbts{3}},-\eta_{\mbts{4}}}\,
A^{J\,(\mbsu{res+pp})}_{(12)\eta_{\mbts{1}},\,(34)\eta_{\mbts{3}}}
(\omega)\,.
\label{ajpp1}
\ee
Propagator
$A^{J\,(\mbsu{res+pp})}_{(12)\eta,\,(34)\eta'}(\omega)$
is a solution of the equation
\be
A^{J\,(\mbsu{res+pp})}_{(12)\eta,\,(34)\eta'} (\omega) =
\tilde{A}^{J\vphantom{(-)}}_{(12)\eta,\,(34)\eta'} (\omega) +
\sum_{\eta''} \sum_{(56)}
\theta^{\vphantom{(+)}}_{(65)}\,
\bar{{\cal K}}^{J\,(\mbsu{res+pp})}_{(12)\eta,\,(56)\eta''} (\omega)\,
A^{J\,(\mbsu{res+pp})}_{(56)\eta'',\,(34)\eta'} (\omega)\,,
\label{ajpp2}
\ee
where
\be
\tilde{A}^{J\vphantom{(-)}}_{(12)\eta,\,(34)\eta'} (\omega) =
- \frac{\eta\,\delta^{\vphantom{(+)}}_{\eta,\eta'}\bigl[
\delta^{\vphantom{(+)}}_{(13)}\,
\delta^{\vphantom{(+)}}_{(24)} +
(-1)^{J+l_{\mbts{1}}-l_{\mbts{2}}+j_{\mbts{1}}-j_{\mbts{2}}}\,
\delta^{\vphantom{(+)}}_{(14)}\,
\delta^{\vphantom{(+)}}_{(23)}\bigr]}
{2\,\bigl(\omega - \eta\,\bigl[E_{(1)} + E_{(2)}\bigr]\bigr)}\,,
\label{ajpp3}
\ee
\vspace{1ex}
\be
\bar{{\cal K}}^{J\,(\mbsu{res+pp})}_{(12)\eta,\,(34)\eta'} (\omega) =
\frac{\eta\,\bigl[
\bar{\Phi}^{J\,(\mbsu{res+pp})}_{(12)\eta,\,(34)\eta'} (\omega) +
(-1)^{J+l_{\mbts{1}}-l_{\mbts{2}}+j_{\mbts{1}}-j_{\mbts{2}}}\,
\bar{\Phi}^{J\,(\mbsu{res+pp})}_{(21)\eta,\,(34)\eta'} (\omega)\bigr]}
{\omega - \eta\,\bigl[E_{(1)} + E_{(2)}\bigr]}\,,
\label{kjrpp}
\ee
\vspace{1ex}
\be
\bar{\Phi}^{J\,(\mbsu{res+pp})}_{(12)\eta,\,(34)\eta'} (\omega) =
\sum_{\eta_{\mbts{1}}\eta_{\mbts{2}}\eta_{\mbts{3}}\eta_{\mbts{4}}}
\delta^{\vphantom{(+)}}_{\eta_{\mbts{1}},\,\eta\vphantom{\eta'}}\,
\delta^{\vphantom{(+)}}_{\eta_{\mbts{2}},-\eta\vphantom{\eta'}}\,
\delta^{\vphantom{(+)}}_{\eta_{\mbts{3}},\,\eta'}\,
\delta^{\vphantom{(+)}}_{\eta_{\mbts{4}},-\eta'}\,
\bar{\Phi}^{J\,(\mbsu{res+pp})}_{[12,34]} (\omega)\,,
\label{phijrpp}
\ee
\vspace{1ex}
\be
\bar{\Phi}^{J\,(\mbsu{res+pp})}_{[12,34]} (\omega) =
\Phi^{J\,(\mbsu{res})}_{[12,34]} (\omega) -
\Phi^{J\,(\mbsu{res})}_{[12,34]} (0) +
{\cal F}^{J(\mbsu{pp})}_{[12,34]}\,.
\label{phipp}
\ee
The order-bounding factors $\,\theta^{\vphantom{(+)}}_{(21)}$
in Eq.~(\ref{ajpp2}) are defined as follows:
$\,\theta^{\vphantom{(+)}}_{(21)} = 1\,$
if the ordinal number of the state $(1)$ is lesser than
the number of $(2)$ $\,[(1) < (2)]$,
$\,\theta^{\vphantom{(+)}}_{(21)} = \frac{1}{2}\,$ if $\,(1) = (2)$,
$\,\theta^{\vphantom{(+)}}_{(21)} = 0\,$ if $\,(1) > (2)$.
The interaction amplitude $\Phi^{J\,(\mbsu{res})}_{[12,34]} (\omega)$
responsible for the QPC is defined by Eq.~(B14) of Ref.~\cite{QTBA2}.
Introducing the notation
\be
{\cal F}^{J(\mbsu{pp})}_{(12)\eta,\,(34)\eta'} =
\sum_{\eta_{\mbts{1}}\eta_{\mbts{2}}\eta_{\mbts{3}}\eta_{\mbts{4}}}
\delta^{\vphantom{(+)}}_{\eta_{\mbts{1}},\,\eta\vphantom{\eta'}}\,
\delta^{\vphantom{(+)}}_{\eta_{\mbts{2}},-\eta\vphantom{\eta'}}\,
\delta^{\vphantom{(+)}}_{\eta_{\mbts{3}},\,\eta'}\,
\delta^{\vphantom{(+)}}_{\eta_{\mbts{4}},-\eta'}\,
{\cal F}^{J(\mbsu{pp})}_{[12,34]}
\label{fpp1}
\ee
and using Eqs. (C2)--(C4) of Ref.~\cite{QTBA2} we obtain the following
ansatz for this quantity
\bea
&&{\cal F}^{J(\mbsu{pp})}_{(12)\eta,\,(34)\eta'} =
\delta_{q_{\mbts{1}},\,q_{\mbts{2}}}\,
\delta_{q_{\mbts{3}},\,q_{\mbts{4}}}\,
\delta_{q_{\mbts{1}},\,q_{\mbts{3}}}\,
\frac{1}{2J+1}
\langle j_2 l_2 ||\,T_{JJ\,0}\,|| j_1 l_1 \rangle
\langle j_4 l_4 ||\,T_{JJ\,0}\,|| j_3 l_3 \rangle
\nonumber\\
&&\times
\bigl[ \delta^{\vphantom{(J)}}_{\eta, \eta'}\,
(u^{\vphantom{(J)}}_{(1)}u^{\vphantom{(J)}}_{(2)}
 u^{\vphantom{(J)}}_{(3)}u^{\vphantom{(J)}}_{(4)}
+v^{\vphantom{(J)}}_{(1)}v^{\vphantom{(J)}}_{(2)}
 v^{\vphantom{(J)}}_{(3)}v^{\vphantom{(J)}}_{(4)})
- \delta^{\vphantom{(J)}}_{\eta, -\eta'}\,
(u^{\vphantom{(J)}}_{(1)}u^{\vphantom{(J)}}_{(2)}
 v^{\vphantom{(J)}}_{(3)}v^{\vphantom{(J)}}_{(4)}
+v^{\vphantom{(J)}}_{(1)}v^{\vphantom{(J)}}_{(2)}
 u^{\vphantom{(J)}}_{(3)}u^{\vphantom{(J)}}_{(4)})\bigr]
\nonumber\\
&&\times\,
\int_0^{\infty}dr\,r^2\,
R^{\vphantom{(J)}}_{(1)}(r)\,R^{\vphantom{(J)}}_{(2)}(r)\,
R^{\vphantom{(J)}}_{(3)}(r)\,R^{\vphantom{(J)}}_{(4)}(r)\,
{\cal F}^{\xi}(r)\,.
\label{fpp2}
\eea

Note that the value of
${\cal F}^{J(\mbsu{pp})}_{(12)\eta,\,(34)\eta'}$ in Eq.~(\ref{fpp2})
of the present paper differs from the corresponding value
derived from Eqs. (C2) and (C3) of Ref.~\cite{QTBA2} by a factor
$\frac{1}{2}$ due to the shorthand summation used in Eq.~(C1)
[$(3)\leqslant (4)$].
In addition, in the case $J=0$ one should set:
$
\,{\cal F}^{J(\mbsu{pp})}_{(12)\eta,\,(34)\eta'} =
\delta^{\vphantom{(+)}}_{(12)}\,\delta^{\vphantom{(+)}}_{(34)}\;
{\cal F}^{J(\mbsu{pp})}_{(11)\eta,\,(33)\eta'}
$
in order to obtain consistency with the gap equation (A25)
of Ref.~\cite{QTBA2} written in the diagonal approximation.
Note that this method is applicable both in the QTBA and in the QRPA.
In the latter case the amplitudes $\Phi^{J(\mbsu{res})}$
in Eq.~(\ref{phipp}) are set to be equal to zero.

\newpage
\end{document}